# ARCHETYPES FOR REPRESENTING DATA ABOUT THE BRAZILIAN PUBLIC HOSPITAL INFORMATION SYSTEM AND OUTPATIENT HIGH COMPLEXITY PROCEDURES SYSTEM

Sergio Miranda Freire[1], Luciana Tricai Cavalini[1], Douglas Teodoro[1], Erik Sundvall[2,3]

[1] Departamento de Tecnologias da Informação e Educação em Saúde/Universidade do Estado do Rio de Janeiro, Rio de Janeiro, Brasil
[2] Department of Biomedical Engineering, Linköping University, Linköping, Sweden
[3] County Council of Östergötland, S-581 91, Linköping, Sweden

**Resumo**: O Ministério da Saúde definiu, por meio da portaria nº 2073 de 29/11/2011, que o modelo openEHR é o padrão para os registros eletrônicos de saúde no Brasil. Este trabalho apresenta um conjunto de arquétipos para representar os principais dados do Sistema de Informação Hospitalar e do módulo de alta complexidade do Sistema de Informação Ambulatorial do SUS, este último a partir de 2008. A base pública do Clinical Knowledge Manager (CKM), da fundação openEHR, foi consultada para selecionar os arquétipos que seriam utilizados para representar os dados dos sistemas mencionados acima. Para diversos conceitos, foi necessário especializar arquétipos do CKM, ou projetar novos arquétipos. Um total de 22 arquétipos foram utilizados, sendo 8 novos, 5 especializados e 9 originais do CKM. Estes arquétipos podem ser utilizados tanto para troca de informações entre sistemas, como também para geração de bases para testes de sistemas baseados no openEHR.

**Palavras-chave:** Sistemas de Informação em Saúde, Modelagem de Dados, Arquétipos.

*Abstract*: The Brazilian Ministry of Health has selected the openEHR model as a standard for electronic health record systems. This paper presents a set of archetypes to represent the main data from the Brazilian Public Hospital Information System and the High Complexity Procedures Module of the Brazilian public Outpatient Health Information System. The archetypes from the public openEHR Clinical Knowledge Manager (CKM), were examined in order to select archetypes that could be used to represent the data of the above mentioned systems. For several concepts, it was necessary to specialize the CKM archetypes, or design new ones. A total of 22 archetypes were used: 8 new, 5 specialized and 9 reused from CKM. This set of archetypes can be used not only for information exchange, but also for generating a big anonymized dataset for testing openEHR-based systems.

## Introduction

The openEHR[1] model comprises a set of specifications designed to improve the maintainability and semantic interoperability of electronic health record (EHR) systems. It is based on a multi-level modelling approach[2], where the basic level consists of a *reference model* that describes generic building blocks and structures of the EHR. The (clinical) domain concepts, e.g., body weight, are expressed by composing and constraining building blocks from the reference model into logical tree structures called *archetypes*. If archetypes are shared and used (preferably internationally) then possibilities of interoperable data exchange (e.g. communicating health record extracts between EHR systems), shared Computerized Decision Support Systems development and multicenter queries/studies are improved. Another level, called *templates*, allows archetypes to be combined and customized in order to design use-case specific forms, messages, etc.

The Brazilian department of health informatics of the Ministry of Health (DATASUS) has developed several public health information systems. Two of them, namely the Hospital Information System (HIS)[4] and the Outpatient Information System (OIS), were designed for claim reimbursement purposes, but they can also be used for service management. The data stored in these systems are produced in public and private healthcare providers and communicated to DATASUS using electronic forms based on a standard defined centrally by DATASUS. Anonymized data is then openly published.

In 2011 the Brazilian Ministry of Health established openEHR as a standard model for EHR systems[3]. A use case that may occur in the future is the one where an openEHR-based EHR system need to communicate extracts containing data about HIS or OIS claims to a system that accepts openEHR extracts and process these claims. In such scenario, it is necessary to have a set of archetypes that represents this data.

The objective of this paper is to collect and design a set of archetypes to represent data from the HIS and High Complexity Procedures Module (HCPM)[5] of the OIS systems of DATASUS.

**Methods**

The manuals and data dictionaries for the HIS and HCPM were used in order to select and design archetypes. The HCPM underwent a change in the data structure in 2008, so this study dealt with the data structure from 2008 onwards. The HCPM contains several tables, each containing attributes related to a specific procedure or health care specialty, namely: bariatric surgery, chemotherapy, radiotherapy, nephrology, medicines, and miscellaneous. The tables in HCPM and HIS share many attributes (although sometimes with different names) in common, such as the identification of the patients, healthcare professional and healthcare unit identification, procedure code, and claim identification. Except for the procedure code and healthcare unit identification, the attributes in common were not mapped to archetypes since this study were only interested in those attributes contained in the files that are publicly available through the DATASUS website[6] and that did not identify the healthcare professional.

Due to lack of space, we will show here only a sample of the attributes of the HIS and HCPM systems, grouped in the following categories:
- Procedures: performed procedure, reason for treatment
- Diagnosis: main diagnosis, secondary diagnosis, associated causes, date of diagnosis, metastasis indicator, tumour stage, date of pathological identification
- Laboratory Tests: diuresis, glucose, albumin, HIV, urea reduction rate
- Physical Measurements: body weight, height, body mass Index
- Administrative Data
    - High Complexity Procedures: reason for encounter, healthcare unit, patient age, date of first dialysis, indication of transplantation, duration of treatment (months), date of begin of treatment chemotherapy, date of begin of treatment radiotherapy, claim reason, death indicator
    - Hospitalization: admission date, ward, bed, type of hospitalization, specialty, healthcare unit, discharge date
- Demographic Data: birth date, gender, race, ethnic group, death date, nationality, level of education

To represent the content of the categorical data elements, DATASUS uses a mix of local and established coding systems. More specifically, three established coding systems are used: i) ICD 10 is used to encode diagnoses; ii) CNES, a Brazilian healthcare provider register, is used to encode healthcare settings information; and iii) SIGTAP, a Brazilian procedure, medication, and materials terminology, is used to encode the performed procedures. For the other categorical data elements, such as gender and reason for encounter, there is one flat definition file, where each code is associated to a label in Portuguese. For example, for gender, the codes F, M and 0 represent the concepts female, male and not required, respectively.

The openEHR Clinical Knowledge Manager[7] is a publicly available repository of archetypes designed according to the openEHR archetype object model. The archetypes in CKM are in different stages of the editing process and may be consulted and downloaded. We tried to reuse CKM archetypes as much as possible. When no archetype was found in the CKM that accommodated a specific attribute, we tried to find an archetype that could be specialized in order to include the attribute. Only when no specialization was considered appropriate we designed a new archetype using an archetype editor[8]. Templates were designed for the HIS and for each table of the HCPM using a template editor[9].

**Results and Discussion**

Table 1 shows the list of archetypes that were either reused, specialized or designed from scratch to represent the data on the HIS and HCPM systems, as well the included attributes in the DATASUS' systems.

Table 1: List of archetypes, attributes from the DATASUS' systems and origin of the archetype.

| Archetype | Attributes | origin of archetype |
|---|---|---|
| openEHR-EHR-ACTION.procedure-sus.v1 | procedure, reason for procedure, time, vascular access, irradiated area, fields/insertions | specialized |
| openEHR-EHR-ADMIN_ENTRY.admission.v1 | admission type, hospital service, admit date/time, state/province, healthcare unit | CKM |
| openEHR-EHR-ADMIN_ENTRY.demographic_data.v1 | gender, race, ethnic group, nationality, birth date, educational level | new |
| openEHR-EHR-ADMIN_ENTRY.high_complexity_procedures_sus.v1 | issue date, reason for encounter, healthcare unit, age, state, duration of treatment, schema, date of beginning of chemotherapy, date of beginning of radiotherapy, number of transplantations, indicator of transplantation, enrolled for transplantation, abdominal ultrasonography, venous fistula amount, date of first dialysis | new |
| openEHR-EHR-ADMIN_ENTRY.hospitalization_authorization.v1 | Intensive Care Unit – total number of days, issue date | new |
| openEHR-EHR-ADMIN_ENTRY.patient_discharge.v1 | date of discharge, reason for discharge, death indicator, hospital infection, claim reason | new |
| openEHR-EHR-CLUSTER.fluid.v1 | volume | CKM |
| openEHR-EHR-CLUSTER.tnm_staging-sus.v1 | Topography, date of pathological identification, clinical staging, histopathological grading | specialized |
| openEHR-EHR-EVALUATION.bariatric_surgery_evaluation.v1 | follow-up in months, Baros score, Baros table | new |

Table 1 – continued.

| Archetype | Attributes | origin of archetype |
|---|---|---|
| openEHR-EHR-EVALUATION.problem-diagnosis-sus.v1 | main diagnosis, secondary diagnosis, associated causes, regional linphonodes | specialized |
| openEHR-EHR-INSTRUCTION.request-procedure.v1 | requested procedure | CKM |
| openEHR-EHR-OBSERVATION.body_mass_index.v1 | body mass index | CKM |
| openEHR-EHR-OBSERVATION.body_weight.v1 | weight | CKM |
| openEHR-EHR-OBSERVATION.height.v1 | height | CKM |
| openEHR-EHR-OBSERVATION.lab_test-antigen_antibody_sus.v1 | HbsAg, HIV, HIC - antibodies | specialized |
| openEHR-EHR-OBSERVATION.lab_test-blood_glucose.v1 | glucose | CKM |
| openEHR-EHR-OBSERVATION.lab_test-hba1c.v1 | HB | CKM |
| openEHR-EHR-OBSERVATION.lab_test-liver_function.v1 | albumine | CKM |
| openEHR-EHR-OBSERVATION.lab_test-urea_and_electrolytes-sus.v1 | urea reduction rate | specialized |
| openEHR-EHR-COMPOSITION.demographic_data.v1 | | new |
| openEHR-EHR-COMPOSITION.hospitalisation.v1 | | new |
| openEHR-EHR-COMPOSITION.outpatient_high_complex_procedures.v1 | | new |

A total of 22 archetypes were used: 8 new, 5 specialized and 9 reused from CKM.. A spreadsheet describes the mapping of each variable in the original tables to the respective element in an archetype. The spreadsheet, the archetypes that are specialized or new and the templates may all be obtained by contacting the authors.

Although the majority of the archetypes in the CKM are still in the draft stage of their lifecycle, the clinical attributes in HIS and HCPM, with the exception of the ones associated with the bariatric surgery evaluation, are found in the CKM archetypes or there are archetypes that can be specialized in order to include them. Apart from the bariatric surgery archetype, the new archetypes were created either for accommodating administrative attributes, several of them specific to the Brazilian health information systems, or for representing compositions designed to create templates.

After excluding the patient and healthcare professional identifying attributes, there were few demographic attributes left. Therefore, instead of using the CKM demographic archetypes to map these attributes, we preferred to include them in an ADMIN_ENTRY archetype. This rationale is also used in some clinical archetypes in CKM that include demographic data. This modeling approach is not optimal in all settings and is thus of course open for discussions and refinements; nevertheless the study shows that it is relatively easy to map the Brazilian public health information system models to archetypes.

The data contained in these information systems includes the main types of variables present in EHRs: textual, temporal, coded and numeric variables. In terms of the openEHR data types, they are distributed as follows: real numerical quantities (DV_QUANTITY) - 6, boolean (DV_BOOLEAN) – 7, codes (DV_CODED TEXT) – 23, integer quantities (DV_COUNT) – 7, dates (DV_DATE) – 7, date-time (DV_DATE_TIME) – 3, proportions (DV_PROPORTION) – 2, strings (DV_TEXT) – 7.

These archetypes and conversion mappings may be useful in at least two cases: for communication purposes and for building a test database.

In the first case, mainly of Brazilian interest, systems designed according to the openEHR specifications could use these archetypes in order to communicate extracts of their records to systems that process the hospitalizations or high complexity procedures claims in Brazil.

The second, internationally interesting, case is to meet a need that is often expressed in openEHR discussion lists; the need for a big, publicly available, reasonably realistic openEHR compliant dataset, for the purpose of testing and comparing openEHR implementations and their performance. The archetypes presented in this study can be used to generate such dataset by using the public available data in the DATASUS website covering several years of anonymized data from all Brazilian states.

**Conclusions**

A set of archetypes that could be used to map the clinical and most administrative data of the Brazilian Public Hospital Information System and the High Complexity Procedures Module of the Brazilian public Outpatient Health Information System is presented. The clinical archetypes in this set are mainly reused from CKM or are specializations of CKM archetypes. These archetypes can be used not only for information exchange between systems, but also for generating datasets for testing and comparing openEHR-based systems.

**Acknowledgements**

To CNPq (grant 150916/2013-2) and INCT-MACC (grant N° 15/2008 MCT/CNPq/FNDCT/CAPES/FAPEMIG/FAPERJ/FAPESP/INSTITUTOS NACIONAIS DE CIÊNCIA E TECNOLOGIA) for the financial support.

**References**

[1] The openEHR Foundation. Available: http://www.openehr.org. Accessed 10 Jul 2014.
[2] Beale T, Heard S. **Architecture overview**. 2008. Available: http://www.openehr.org/releases/1.0.2/architecture/overview.pdf. Accessed 10 Jul 2014.
[3] Brasil. Ministério da Saúde, Gabinete do Ministro. Portaria nº 2073, de 31 de agosto de 2011- Regulamenta o uso de padrões de interoperabilidade e informação em saúde para sistemas de informação em saúde no âmbito do Sistema Único de Saúde, nos níveis Municipal, Distrital, Estadual e Federal, e para os sistemas privados e do setor de saúde suplementar. Available: http://bvsms.saude.gov.br/bvs/saudelegis/gm/2011/prt2073_31_08_2011.html. Accessed 10 Jul 2014.
[4] Brasil. Ministério da Saúde, DATASUS. Sistema de Informação Hospitalar Descentralizado. Dicionário de dados. Disponível http://www2.datasus.gov.br/SIHD/manuais. Accessed 10 Jul 2014.
[5] Brasil. Ministério da Saúde, DATASUS. Disseminação de Informações do Sistema de Informações Ambulatoriais do SUS (SIASUS), Informe Técnico 2014-01; 2014; Available: ftp://ftp.datasus.gov.br/dissemin/publicos/SIASUS/200801_/Doc/IT_SIASUS_1401.pdf. Accessed 10 Jul 2014.

[6] DATASUS. Portal de Saúde. Transferência de Arquivos. Available: http://www2.datasus.gov.br/DATASUS/index.php?area=0901&item=1&acao=22&pad=31655. Access in jul 10, 2014.
[7] Ocean Informatics. Clinical Knowledge Manager (CKM). Available: http://www.openehr.org/ckm/. Accessed 10 Jul 2014.
[8] Ocean Informatics. Ocean Archetype Editor. Available: http://www.openehr.org/svn/knowledge_tools_dotnet/TRUNK/ArchetypeEditor/Help/index.html. Accessed 10 Jul 2014..
[9] Ocean Informatics: Ocean Template Designer. Available: http://wiki.oceaninformatics.com/confluence/display/TTL/Template+Designer+Releases. Accessed 10 Jul 2014..

**Contact**

Sergio Miranda Freire, associate professor
Departamento de Tecnologias da Informação e Educação em Saúde, Universidade do Estado do Rio de Janeiro
Av. Prof. Manuel de Abreu, 444/2º andar, Vila Isabel Rio de Janeiro, Brasil - 20550-170
e-mail: sergio@lampada.uerj.br
Phone: (021) 2868-8378